\documentclass[10pt,conference]{IEEEtran}
\IEEEoverridecommandlockouts
\pdfoutput=1
\usepackage{cite}
\usepackage{amsmath,amssymb,amsfonts}
\usepackage{algorithmic}
\usepackage{graphicx}
\usepackage{textcomp}
\usepackage{xcolor}

\def\BibTeX{{\rm B\kern-.05em{\sc i\kern-.025em b}\kern-.08em
    T\kern-.1667em\lower.7ex\hbox{E}\kern-.125emX}}

\begin{document}

\title{Simultaneous or Sequential Training? How Speech Representations Cooperate in a Multi-Task Self-Supervised Learning System
{\footnotesize \textsuperscript{*}}
\thanks{This research was funded by Academy of Finland grants no. 314602 and 345365, and Kone Foundation grant L-SCALE.}
}

\author{\IEEEauthorblockN{1\textsuperscript{st} Khazar Khorrami}
\IEEEauthorblockA{\textit{Unit of Computing Sciences} \\
\textit{Tampere University}\\
Tampere, Finland \\
firstname.surname@tuni.fi}
\and
\IEEEauthorblockN{2\textsuperscript{nd} María Andrea Cruz Blandón}\IEEEauthorblockA{\textit{Unit of Computing Sciences} \\
\textit{Tampere University}\\
Tampere, Finland \\
firstname.surname@tuni.fi}
\and
\IEEEauthorblockN{3\textsuperscript{nd}Tuomas Virtanen}
\IEEEauthorblockA{\textit{Unit of Computing Sciences} \\
\textit{Tampere University}\\
Tampere, Finland \\
firstname.surname@tuni.fi}
\and
\IEEEauthorblockN{4\textsuperscript{rd} Okko Räsänen}
\IEEEauthorblockA{\textit{Unit of Computing Sciences} \\
\textit{Tampere University}\\
Tampere, Finland \\
firstname.surname@tuni.fi}
}

\maketitle

\begin{abstract}
Speech representation learning with self-supervised algorithms has resulted in notable performance boosts in many downstream tasks. Recent work combined self-supervised learning (SSL) and visually grounded speech (VGS) processing mechanisms for representation learning. The joint training with SSL and VGS mechanisms provides the opportunity to utilize both unlabeled speech and speech-related visual information based on data availability. This has shown to enhance the quality of learned representations, especially at encoding semantic- and lexical-level knowledge. In this work, we further study the joint optimization of wav2vec 2.0-based SSL and transformer-based VGS as a multi-task learning system. We explore a set of training scenarios to understand how speech representations are shared or transferred between the two tasks, and what is the optimal training strategy for cross-modal semantic retrieval and phoneme discrimination performance. As a result, we find that sequential training with wav2vec 2.0 first and VGS next provides higher performance on audio-visual retrieval compared to simultaneous optimization of both learning mechanisms. However, the parallel SSL-VGS training reduces the effects of catastrophic forgetting when switching between optimization criteria. Moreover, the results suggest that phonemic representations learned through the VGS mechanism may generalize better across datasets compared to those learned with SSL. 
\end{abstract}

\begin{IEEEkeywords}
speech representation learning, visually grounded speech, multi-task learning, multi-modal neural networks
\end{IEEEkeywords}




\section{Introduction and related works}\label{Intro}

Visually grounded speech (VGS) processing refers to algorithms that learn correspondences between image and speech data in an unsupervised manner (see \cite{chrupala2022visually} for a review). VGS models are central to the study of autonomous AI systems that could ground their world knowledge to multimodal associations. In addition, they are commonly used to model human infant language learning \cite{khorrami2021can}. The data for training a VGS model comes in the form of images paired with spoken descriptions of the images. In a typical VGS system, speech and image data are processed in parallel neural modules and then mapped together into a shared embedding space, where a similarity score -based contrastive optimization is used for network training (see, e.g., \cite{wang2021understanding}). The system is usually evaluated for its performance on speech-to-image and image-to-speech retrieval tasks (see \cite{chrupala2022visually}).  

Previous research has shown that hidden layers of trained VGS models reflect linguistic information at, e.g., phonemic and lexical levels, showing that the models can be used for (multimodal) speech representation learning \cite{khorrami2021can,chrupala2017representations,alishahi2017encoding,harwath2017learning} . This is similar to unimodal algorithms for self-supervised learning (SSL), such as wav2vec 2.0 \cite{baevski2020wav2vec}, HuBERT \cite{hsu2021hubert}, and CPC \cite{oord2018representation}, that learn useful speech representations using acoustic speech input as the only data. Similar to large-scale language models (e.g., BERT \cite{kenton2019bert}), the speech representations learned through self-supervised models have shown notable performance boosts in many supervised downstream tasks, such as phoneme or emotion recognition \cite{baevski2020wav2vec, vaaras2022analysis, yang2021superb}, thereby having potential especially in low-resource speech tasks.  

Recently, Peng and Harwath \cite{peng2022self, peng2022word} introduced a system that jointly learns speech representations using acoustic-level self-supervised learning and semantic-level visual grounding. The use of two learning mechanisms provides a potential advantage over the individual mechanisms: the system can process a combination of speech-only (unlabeled) data through a SSL block and (weakly labeled) speech-image pairs through a VGS block according to data availability in the two cases. This enables potentially synergetic and flexible learning of speech representations from the two previously established representation learning mechanisms, as several layers of the speech encoder module are shared between the SSL and VGS networks. Using this model, Peng and Harwath \cite{peng2022self} showed that the joint model performs competitively on phonemic task of ZeroSpeech 2021 challenge \cite{dunbar2022self} and SUPERB benchmark \cite{yang2021superb} while also outperforming many models at semantic and lexical tasks. 

The joint VGS and SSL training, as in \cite{peng2022self} and \cite{peng2022word}, can be seen as a multi-task multi-domain system with the capacity for both incremental and simultaneous learning. While catastrophic forgetting is a main challenge in domain-incremental learning, the major problem with the task-incremental learning is to obtain the knowledge that can be transferred across tasks \cite{van2022three}.
However, the potential synergies of a shared encoder for SSL and VGS tasks were not comprehensively studied in the earlier works (\cite{peng2022self,peng2022word}). In addition, most of the experiments by Peng and Harwath used an additional corpus (LibriSpeech \cite{panayotov2015librispeech}) for SSL learning or pre-trained weight initialization, making it difficult to disentangle benefits of mechanism synergies per se from potential benefits of simply having more (and more varied) training data for the SSL. Thereby, it remains unclear in what conditions joint SSL and VGS training facilitates the learning process (e.g., in terms of final representation quality, learning rate, or cross-corpus generalization) compared to what is obtained by the individual mechanisms, and whether benefits of joint training also occur in a case where only the same audio data is available to both learning mechanisms. Also, if the joint training provides learning improvements, what would be the best way to schedule learning with the SSL and VGS mechanisms (e.g., parallel or sequential optimization)? 

In this work we try to answer the above questions by investigating a set of combined SSL and VGS training scenarios using the system from \cite{peng2022self} for speech-to-image semantic mapping and self-supervised acoustic modeling. We study how temporal sequencing of the learning mechanisms affects phonemic and audiovisual semantic learning when both mechanisms have access to the same audio data.




\section{Model description}\label{Model}

\begin{figure}[htbp]
\centerline{\includegraphics[width=1\columnwidth]{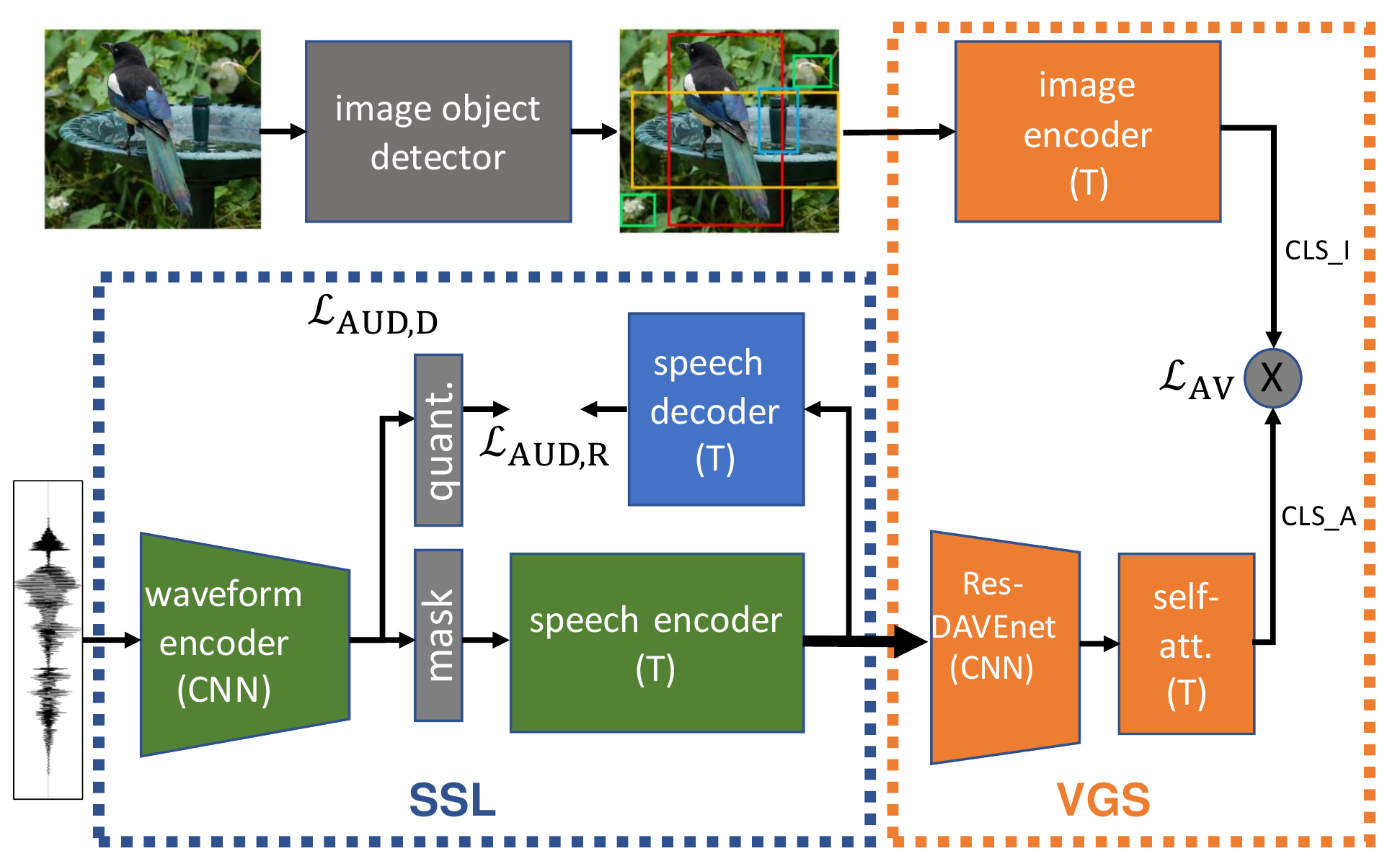}}
\caption{VGS+ as a joint model of VGS and SSL training. The green block is optimized for the both tasks.}
\label{figModel}
\end{figure}

We adopted the FaST-VGS+ model from \cite{peng2022self} with a simplification of using only the "coarse" audio-visual loss of the model for computational feasibility (i.e., the "Fast" transformer version; see \cite{peng2022fast} for more details). The model (Fig. \ref{figModel}) consists of two main mechanisms for speech-image training (a transformer-based VGS model \cite{peng2022fast}) and speech SSL training (masking-based speech acoustic modeling with wav2vec 2.0; \cite{baevski2020wav2vec}). Most of the speech encoder is shared between the VGS and SSL mechanisms and optimized for the both tasks (Fig. \ref{figModel}, green block). 

In the VGS pipeline, the image and speech inputs are processed in parallel branches where the classification (CLS) tokens of the last transformer layers are used as "semantic" speech and image embeddings. These embeddings are compared using cosine similarity score and optimized for similarity (dissimilarity) in case of matching (mismatching) speech-image pairs. The speech-based SSL uses wav2vec 2.0 (from now on: W2V2) network that randomly masks segments of input speech and learns speech representations by predicting the masked sections from other parts of the same utterance. 

The audio waveform encoder shared by SSL and VGS is a 6-layer convolutional neural network (CNN) that maps the input acoustic waveform (in 16 KHz) to embedding of 512-d (calculated every 10 ms). It is followed by an 8-layer transformer block ("speech encoder" in Fig. 1) shared between the two VGS and W2V2 networks, and 4 additional transformer layers dedicated to W2V2 only ("speech decoder"). ResDAVEnet \cite{hsu2019transfer} is stack of convolutional and pooling layers applying the down-sampling in time, and the image encoder is a 6-layer transformer block. The dimension of all transformer layers is 768. 
The VGS network is trained through a masked and marginalized "InfoNCE" loss \cite{ilharco2019large} (here denoted as $loss_{AV}$) as a contrastive learning method that tries to minimize the distance between ground-truth speech-image pairs compared to a set of disctractor random pairs taken from the same training mini-batch. In W2V2, a contrastive masking loss tries to minimize the distance between the masked speech representations and their ground-truth quantized versions compared to a set of distractors coming from the same utterance. Moreover, a diversity loss is used to encourage the equal use of codebook entries at quantization block. 

Following the original work \cite{baevski2020wav2vec}, we combined the two masking ($loss_{AUD,R}$) and diversity ($loss_{AUD,D}$) losses in proportion of 1:0.1, denoting their sum as $loss_{AUD}$. The VGS+ model is trained by combining the $loss_{AV}$ and $loss_{AUD}$ with a coefficient $\alpha$ that controls the emphasis on the two training mechanisms as
\begin{equation}
loss = \alpha  loss_{AV} + (1-\alpha) loss_{AUD} \label{eq1} .
\end{equation}

By varying $\alpha$ at training time, we could manipulate the contribution (and timing) of the auditory and audiovisual learning mechanisms in the overall system training.

\section{Experiments}

 The aim of the experiments was to understand how the SSL and VGS mechanisms interact in different training scenarios with varying emphasis on auditory and audiovisual losses, and what training strategy results in the best representation quality for phonemic discrimination and audiovisual retrieval tasks.

\subsection{Datasets}\label{Data}
We utilized SpokenCOCO (SC) dataset \cite{hsu2021text} as the training data. It comes with 123k images and 5 spoken English captions per image, resulting a total of 742 h of speech. 
We used 118k images for model training and 5k images for testing on semantic retrieval tasks. Phoneme discrimination of the representations was measured on LibriSpeech (LS) dev-clean subset (denoted by C)\cite{panayotov2015librispeech}.

\subsection{Model variations}\label{Variations}
We considered three base model variants: 
1) VGS only ($\alpha = 1 $), 2) W2V2 only ($\alpha = 0 $), and 3) VGS+ with an equal emphasis on both losses ($\alpha = 0.5 $). We also defined six training variations of these by relative scheduling of the basic optimization approaches. In each variation, one base model was used as a pretraining for follow-up training using another variant. We denote the scheduled system as (A, B), where A is the base model used at the pretraining phase and B is the main training phase. As the first scenario, we pretrained with VGS or W2V2 and continued with VGS+ (i.e, (W2V2, VGS+) and (VGS, VGS+)). As the second scenario, we pretrained with each individual model of VGS or W2V2 and then continued with the other model ((W2V2, VGS) and (VGS, W2V2)). And finally as the third scenario, we pretrained with VGS+ and continued with one of the individual models of VGS or W2V2 ((VGS+, W2V2) and (VGS+, VGS)).
\subsection{Evaluation}\label{Evaluation}
Representation learning was evaluated in terms of semantic and phonemic performance scores, as tested separately for the training variations to investigate what is the optimal scenario regarding the measured performance metrics. We also qualitatively analyzed loss curves to understand how the optimization strategy affects the dynamics of the training process.

As the metric for phonemic representations, we used the ABX phonemic discrimination score by \cite{schatz2013evaluating}, also used as one of the primary metrics in ZeroSpeech challenges \cite{dunbar2022self}. ABX is measured in both within- (W) and across-speaker (A) conditions, whether the latter reflects cross-speaker generality of the learned phonemic distinctions (see \cite{schatz2013evaluating}). 

For evaluating the semantic knowledge of the model, we used recall@k metric \cite{hodosh2013framing} to measure speech-to-image and image-to-speech retrieval performance (see \cite{chrupala2022visually} for more details on evaluation of semantic retrieval tasks in VGS models.)

\subsection{Implementation Details}\label{ID}
The base models were trained for 70 epochs, as based on saturation of semantic retrieval performance in pilot experiments. For the scheduling scenarios, the base model trained for 20 epochs was used as the initialization for another 50 epochs with the other loss weighting configuration.
We used Adam optimizer with an initial learning rate of 10-4, a warm-up fraction of 0.1, and then a linear decay towards the end of the training. The optimizer state was reset after the pretraining process. For the semantic retrieval score, we used classification tokens of the speech and the image embedding layers. We report the recall@10 on 25k test pairs (5k images each paired with 5 spoken captions) at the final epoch of the training. For measuring the ABX phoneme discrimination error, we saved the model every 5 epochs and measured the ABX-error for representations of all the 12 intermediate layers of the speech encoder and decoder blocks (cf. Fig 1).  




\section{Results and Discussion}

\subsection{Semantic retrieval }

For semantic retrieval performance, we measured and compared recall@1 and recall@10 scores for both speech-to-image and image-to-speech retrieval tasks. Table \ref{tabRecall} shows the results on test data obtained at the end of the training process. 

The base models, VGS and VGS+, have similar performance. This result accords with the previous report \cite{peng2022self} and indicates that the semantic retrieval performance does not benefit from simultaneous optimization of speech encoder representations for the auditory SSL (W2V2) task. However, when the W2V2 training precedes VGS training (the (W2V2, VGS+) and (W2V2, VGS) variants), there is a substantial improvement in recall scores. Notably, we observe this improvement when pretraining \textit{on the same data} as the main training, whereas previous improvements have been reported with pretraining on large-amounts of additional speech data not used for the VGS task (LibriSpeech in \cite{peng2022fast, peng2022self}). We also tested and observed that using both LibriSpeech and SpokenCOCO at the pretraining step does not improve the retrieval performance of the (W2V2, VGS) and (W2V2, VGS+) variants above what is gained by using either of the datasets. 
Thus, the improvement is mainly the result of self-supervised initialization, not from using more speech data (see also the results from different data settings in \cite{peng2022self}). We also tested if the semantic scores can be further improved by incorporating more epochs to the W2V2 pretraining phase. However, this did not improve the recall scores from those obtained by the initial setting of 20 epochs. 
\begin{table}[htbp]
\caption{Semantic retrieval results (recall at 1 and 10) for the base and the scheduled systems obtained at the end of training.}
\begin{center}
\begin{tabular}{|l|c|c|c|c|}
\hline
 & \multicolumn{2}{|c|}{\textbf{speech-to-image}} & \multicolumn{2}{|c|}{\textbf{image-to-speech}} \\
\cline{2-5} 
 & \textbf{\textit{r1}}& \textbf{\textit{r10}} &  \textbf{\textit{r1}}& \textbf{\textit{r10}} \\
\hline
\textbf{Base models} & \multicolumn{4}{|c|}{} \\
\hline
W2V2 & 0.0  & 0.2 & 0.0 & 0.2  \\
\hline
VGS &  28.4 & 69.4 &  40.1 & 81.5  \\
\hline
VGS+ & 29.0  & 70.1 & 40.1  & 81.8 \\ 
\hline
\textbf{Scheduled trainings} & \multicolumn{4}{|c|}{} \\
\hline
(W2V2, VGS+) & \textbf{32.0}  & \textbf{74.7}  & \textbf{44.4} & \textbf{85.0}   \\
\hline
(VGS, VGS+) & 27.9  & 68.8 & 39.3  & 81.3  \\
\hline
(W2V2, VGS) & \textbf{32.3}   & \textbf{74.9} & \textbf{44.8}  & \textbf{84.8}  \\
\hline
(VGS, W2V2) & 0.0  & 0.2 & 0.0 & 0.2  \\
\hline
(VGS+, W2V2) & 14.5   & 52.0 & 7.5 & 32.5  \\
\hline
(VGS+, VGS) & 28.8 &  70.0 & 39.8  & 81.5 \\
\hline
\hline
FaST-VGS\textsubscript{CO} (Pre LS) \cite{peng2022fast} & 31.8 & 75.0 & 42.5 & 84.9  \\
\hline
\end{tabular}
\label{tabRecall}
\end{center}
\end{table}

The reason why the same retrieval performance gain is not achieved by simultaneous training from scratch (i.e., VGS+) could be because the VGS loss might be easier to optimize compared to the W2V2 training. This may cause VGS to dominate the training process, resulting in a less optimal overall solution for the W2V2 (see Fig. \ref{figLoss} and the discussion at comparing the two losses curves). 

Finally, catastrophic forgetting quickly results in chance-level recall score when switching from VGS pretraining to W2V2 training. In contrast, the performance remains fairly stable with (VGS+, W2V2). This suggests that synergistic representations between SSL and VGS are possible for audiovisual learning, but require the presence of both mechanisms from the start of the training in order to be robust against later alternation between the two tasks. 

\subsection{Phonemic discrimination }

Table \ref{tabABX} shows the ABX results obtained for the different training variants. In general the VGS outperforms the W2V2 training in all tested variants with the best ABX score obtained at (VGS+, VGS) and then VGS. Bearing in mind that all models are trained on SC that is different domain data from LS used at ABX test, this result suggests that audiovisual training provides phonemic representations that better generalize across data domains. In contrast, pretraining with W2V2 prior to VGS or VGS+ training does not result in equally good cross-dataset generalization in ABX. This is a highly relevant finding that requires further research, considering that SSL models tend to suffer from domain-mismatch problems (\cite{hsu2021robust}). In addition, previous work has not reported any performance benefits for ABX from the use of visual data, but, except for \cite{peng2022self}, all these studies have used LS data for both SSL training and ABX testing (see a comparison in \cite{peng2022self}).

\begin{table}[htbp]
\caption{ABX error measured  within speaker (W-C) and across speakers (A-C) on LibriSpeech (LS) dev-clean set. Chance level is 50.}
\begin{center}
\begin{tabular}{|l|c|c|}
\hline
\textbf{Model} & W-C (layer) & A-C (layer) \\
\hline
\textbf{Base models} & \multicolumn{2}{|c|}{} \\
\hline
W2V2 & 7.05 (3) & 9.53 (4)   \\
\hline
VGS & \textbf{4.82 (4)} & \textbf{6.44 (4)}   \\
\hline
VGS+ & 5.09 (8) & 6.81 (8)   \\
\hline
\textbf{Scheduled trainings} & \multicolumn{2}{|c|}{} \\
\hline
(W2V2, VGS+) & 6.95 (2) & 8.72 (5)  \\
\hline
(VGS, VGS+) & 5.49 (7) & 7.55 (8)  \\
\hline
(W2V2, VGS) & 6.74 (2) & 8.75 (2)  \\
\hline
(VGS, W2V2) & 8.4 (1) & 11.54 (2)  \\
\hline
(VGS+, W2V2) & 6.23 (8) & 8.35 (8)  \\
\hline
 (VGS+, VGS) & \textbf{4.50 (9)} & \textbf{6.02 (9)}   \\
\hline
\hline
FaST-VGS+ (LS+SC) \cite{peng2022self} & 4.24 & 5.08  \\
\hline
\end{tabular}
\label{tabABX}
\end{center}
\end{table}
 
\vspace{-12 pt}
\begin{figure}[htbp]
\centerline{\includegraphics[width=1\columnwidth]{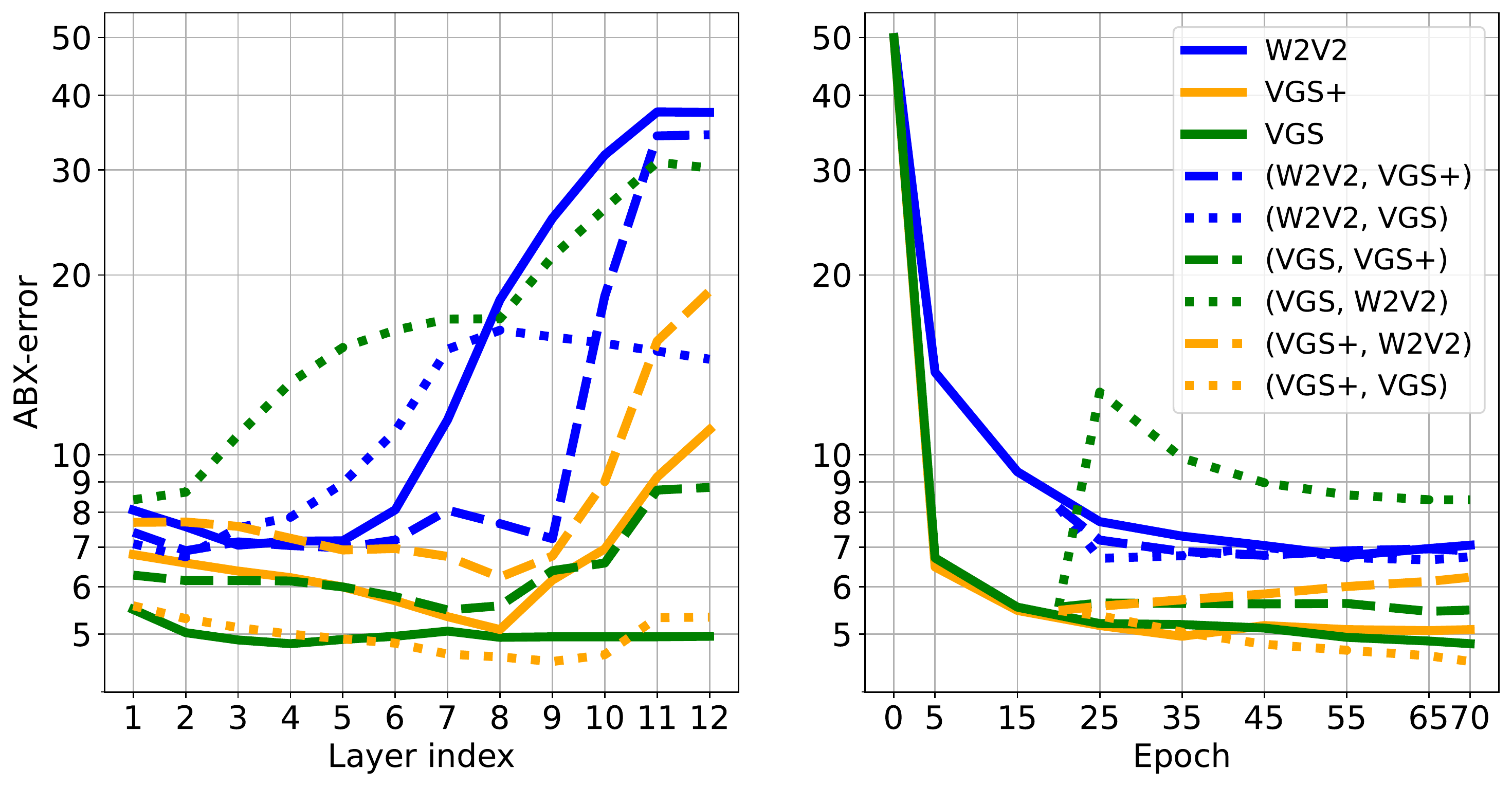}}
\caption{ABX error (W-C) for the tested training variants. Left: as a function of speech encoder (no. 1--8) and decoder (9--12) layer after the full 70 training epochs. Right: best layer score across the training epochs. Note the logarithmic scale of the y-axes.}
\label{figABX-LE}
\end{figure}

Fig. \ref{figABX-LE} (left) shows (W-C) ABX error of the different hidden layers of the transformer-based speech encoder and decoder. The first layers of the encoder perform comparably in all tested variants. In the best performing variant of (VGS+, VGS) the score improves slightly for the deeper encoder layers whereas in VGS all hidden layers perform similarly well in the task. Overall, for the variants having VGS+ at one of the training phases, the error decreases with increasing layer depth in the encoder and increases again for deeper layers of the speech decoder. A similar performance pattern across layers is observed when W2V2 is trained on LS data \cite{peng2022self}.

Fig. \ref{figABX-LE} (right) illustrates the lowest (W-C) ABX error (among layers) during the training epochs. For all base models, the ABX error improves monotonically during the training. For the scheduled versions, the ABX behavior at transitions between the losses is always smooth when the initial training phase is the acoustic-level W2V2 (blue lines). In contrast, in (VGS, W2V2), and then with a slighter rate at (VGS+, W2V2), the error increases substantially when shifting from VGS/VGS+ to W2V2 training. The behavior of the ABX score at transition points suggest that the phonemic representations learned from acoustics can be further transferred to semantic tasks, but the representations learned during the semantic optimization cannot be initially adopted to acoustic learning. However, having in mind that all our variants are trained on SC data, i.e., different domain from the LS test data, further investigation is required to distinguish the effect of domain change from the sheer role of the learning task.

\subsection{Training loss analysis}

\begin{figure}[htbp]
\centerline{\includegraphics[width=1\columnwidth]{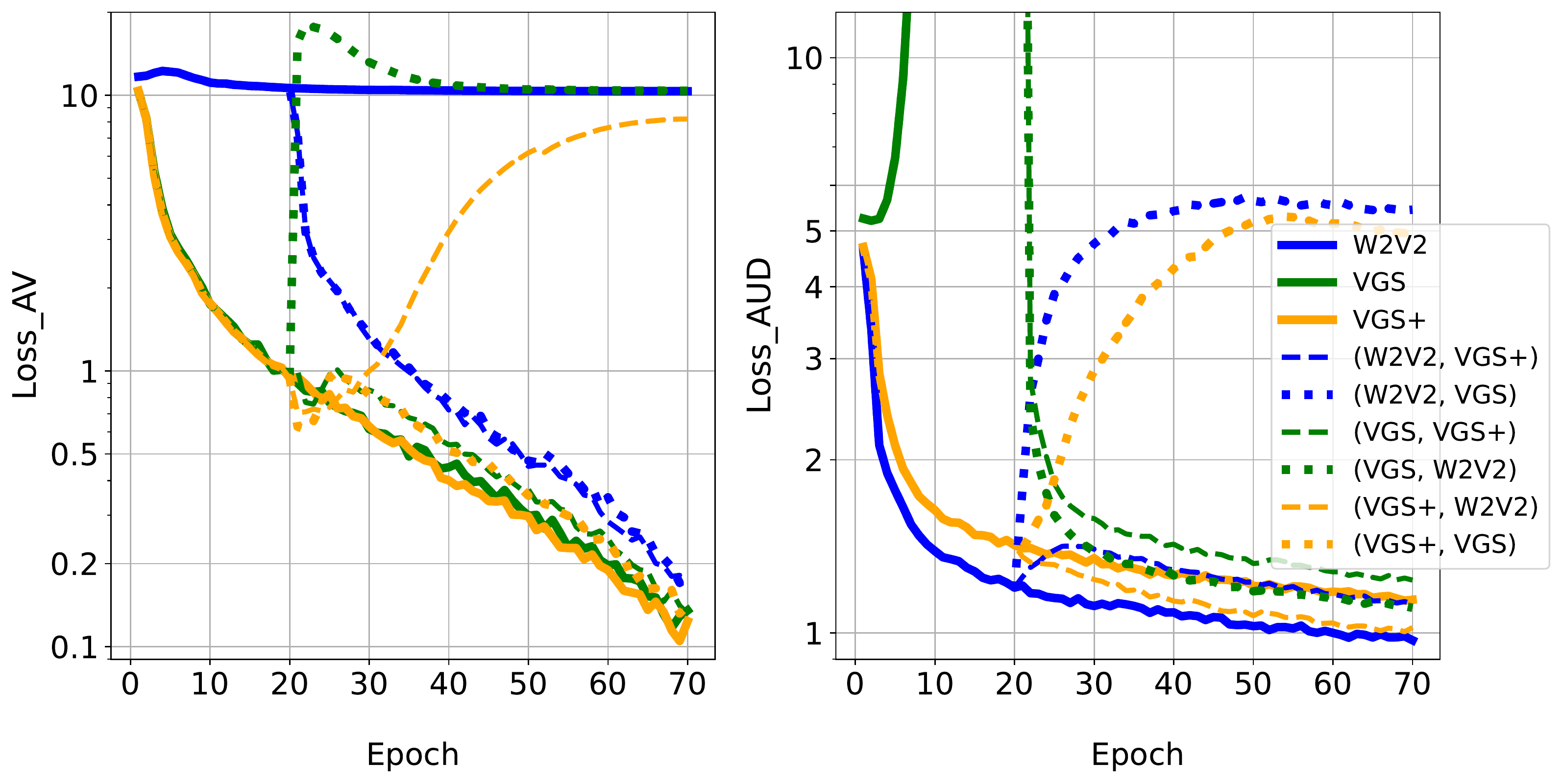}}
\caption{Training loss curves for all tested model variants (log-scale): Left: $loss_{AV} $. Right: $loss_{AUD} $. The graphs show full training (70 epochs) for the base models (solid lines) and the combinations of 20 epochs of pretraining and 50 epochs of main training for the training schedule variants.}
\label{figLoss}
\end{figure}

Fig. \ref{figLoss} shows the loss curves of $loss_{AV}$ (left) and $loss_{AUD}$ (right) for the base and the scheduled models. Although the general range of the two $loss_{AV}$ and $loss_{AUD}$ curves are very close, in overall $loss_{AV} $ decays with steeper slope especially at the later epochs. Comparing the faster decay rate of the $loss_{AV}$ in VGS to decay of $loss_{AUD}$ in W2V2 suggests that audio-visual semantic mapping is an easier task compared to W2V2 acoustic modeling, and the pattern stays same when the two losses are optimized simultaneously (i.e. in VGS+). 

Furthermore, similar to the attitude found at the semantic retrieval scores, catastrophic forgetting also leads to rapid decrease of $loss_{AV} $ to a chance-level when switching from VGS to W2V2. An analogous pattern is observed in the behavior of the $loss_{AUD} $ when switching from W2V2 to VGS. In contrast, the forgetting effect is much milder in the cases where the pretraining phase includes optimization of both losses (i.e.,  VGS+ pretraining). For example, $loss_{AV} $ in (VGS+, W2V2) tolerates the absence of audio-visual updates for a few epochs after which it starts to gradually increase.

\section{Conclusions}

This study set out to investigate the coordination between SSL and VGS mechanisms. We tested a number of training scenarios involving the wav2vec 2.0-based SSL and transformer-based VGS models, and studied the performance of the resulting speech representations in semantic cross-modal retrieval and phoneme discrimination tasks.
The results show that simultaneous learning with SSL and VGS mechanisms does not provide performance gains for phonemic or semantic learning compared to the individual mechanisms. However, joint training ensures synergetic representations that are robust against catastrophic forgetting in the individual tasks in follow-up training with just one mechanism. In contrast, acoustic pretraining prior to audiovisual semantic training boosts the performance on the semantic task, even when the SSL-based pretraining takes place on the same dataset.

Notably, our results show that the best phonemic representations, when evaluated in cross-domain conditions, were obtained by visually-grounded learning, and the representations can be further improved if the visual learning is preceded by simultaneous visual and acoustic learning. This is in contrast to previous findings \cite{peng2022self, dunbar2022self}. However, according to our knowledge, this is the first study to compare SSL and VGS when neither of the mechanisms have access to the corpus used in the ABX test for phonemic discrimination. In future, we plan to further investigate if speech representation learning with the help of visual semantics helps to improve generalization across datasets compared to purely SSL-based speech representations.  

\section*{Acknowledgment}
 The authors would like to thank CSC for computational resources and Puyuan Peng for his valuable help with the FaST-VGS+ model.

\bibliographystyle{IEEEtran}
\bibliography{main}

\end{document}